# Snapshot hyperspectral imaging of intracellular lasers

SORAYA CAIXEIRO,[1,4,†] PHILIP WIJESINGHE,[2,†] KISHAN DHOLAKIA,[2,3] AND MALTE C. GATHER[1,2,5]

[1]*Humboldt Centre for Nano- and Biophotonics, Department of Chemistry, University of Cologne, Greinstr. 4-6, 50939 Cologne, Germany*
[2]*Centre of Biophotonics, SUPA, School of Physics and Astronomy, University of St Andrews, North Haugh, St Andrews, Fife, KY16 9SS, UK*
[3]*Centre of Light for Life and School of Biological Sciences, The University of Adelaide, Adelaide, South Australia, Australia*
[4]*soraya.caixeiro@uni-koeln.de*
[5]*malte.gather@uni-koeln.de*
[†]*These authors contributed equally to this work*

**Abstract:** Intracellular lasers are emerging as powerful biosensors for multiplexed tracking and precision sensing of cells and their microenvironment. This sensing capacity is enabled by quantifying their narrow-linewidth emission spectra, which is presently challenging to do at high speeds. In this work, we demonstrate rapid snapshot hyperspectral imaging of intracellular lasers. Using integral field mapping with a microlens array and a diffraction grating, we obtain images of the spatial and spectral intensity distribution from a single camera acquisition. We demonstrate widefield hyperspectral imaging over a 3×3 mm$^2$ field of view and volumetric imaging over 250×250×800 μm$^3$ volumes with a spatial resolution of 5 μm and a spectral resolution of less than 0.8 nm. We evaluate the performance and outline the challenges and strengths of snapshot methods in the context of characterising the emission from intracellular lasers. This method offers new opportunities for a diverse range of applications, including high-throughput and long-term biosensing with intracellular lasers.

Intracellular lasing | laser particles | integral field mapping | whispering gallery mode lasers | biolaser.

## 1. Introduction

Intracellular lasers are micro-to-nanoscale, tissue-integrated lasing particles that enable multiplexed large-scale tracking and precision sensing in biomedicine [1–4]. They are an attractive alternative to conventional luminescent particles, like quantum dots and fluorescent beads, due to their high spectral purity, very narrow emission linewidth, and high output intensities [5–8]. When integrated into tissues and cells, lasing particles can serve as optical barcodes for multiplexed cell tracking in 2D and 3D [3,4,9,10] and can perform sensing of the local microenvironment due to the minute shifts in their spectral mode position upon changes in local refractive index [1,2,10–14]. Recent studies in cardiac tissue and phantoms have shown that lasing particles can be detected at depth through turbid media, and even localised in 3D [1,15].

To date, most studies on intracellular lasers have involved manually addressing individual lasers point-by-point using a spectrometer, i.e., intracellular lasers are excited by an external pump laser and their emission is collected through a microscope objective and imaged onto the entrance slit of a spectrometer. This approach is sequential by nature, limiting the throughput and speed of detection.

Recording the spectrally rich emission from intracellular lasers across multiple cells distributed across a field of view requires hyperspectral imaging, i.e., a technique that measures both the spectral and spatial intensity of a light field. Hyperspectral imaging acquires spatial and spectral information in multidimensional 'datacubes' or 'hypercubes', e.g., of the form: $I(x, y, \lambda)$ [16], where $\lambda$ denotes the wavelength. Higher dimensions may also include depth, $z$, or time $t$. In many cases spectral features of interest are relatively broad, e.g., when hyperspectral imaging is applied to distinguish different fluorescent dyes or light-absorbing species in a biological sample [17–19], or when air-borne and satellite-based hyperspectral imagers are used for environmental monitoring and in agriculture [20]. In these situations, cameras with appropriate filters or multichannel photo-multiplier tubes enable rapid data acquisition and sufficient spectral resolution. However, to take full advantage of the information encoded in the narrow emission spectra of intracellular lasers, a higher spectral resolution is generally required.

One way to achieve high spectral resolution is though point scanning the sample volume, for instance, using galvanometer mirrors or translation stages, and analysing the laser emission with a spectrometer. This typically limits imaging speed to the integration time and the readout rate of the spectrometer camera; ensuring sufficiently high signal-to-noise ratios in the recorded laser spectra can result in scan times on the order of hours [4,9]. There have been few demonstrations of such hyperspectral imaging of intracellular lasers to date. For instance, fast confocal scanning and spectral readout of the emission from intracellular semiconductor disk lasers has allowed for tracking cell migration in a portion of a tumour spheroid (1 × 1 × 0.28 mm$^3$), with the acquisition of each stack taking ~47 mins [4], which is prohibitively slow in many scenarios. Similarly, a point-scanning spectrometer was integrated with fluorescence microscopy and optical coherence tomography for multimodal structural and spectral imaging of nanowire laser-tracked stem cell migration in rabbit eyes [9]. However, because the spectral collection time was a few seconds for each location, 13 lasers were addressed individually by tracking their positions from their fluorescence signal. Methods for faster hyperspectral imaging of intracellular lasers are thus a burgeoning need for applications that require monitoring of biological processes in real-time and to improve throughput in existing applications.

Outside the realm of intracellular lasers, important advances have been made to accelerate hyperspectral imaging [16]. Fundamentally, increases in speed must come from an increase in the total number of available detector elements; for instance, in 'push-broom' spectrometers that disperse a line scan over a 2D area detector [21], or from a more efficient use of the detector elements, for instance, in the coded-aperture snapshot spectral imager (CASSI) that avails itself of sparsity in the spatio-spectral content to reconstruct high quality hypercubes [22]. Particularly powerful speed advances are achievable from a class of hyperspectral imaging termed snapshot hyperspectral imaging (SSHI) [16]. SSHI describes methods that multiplex both spatial and spectral content into one wide-area detector, such that a hypercube can be reconstructed from one acquisition event of a camera, i.e., a 'snapshot'. These methods lead to video-rate imaging, however, multiplexing 3D information onto a 2D detector requires some trade-off in spatial and spectral resolution. Common SSHI methods are based on integral field mapping (IFM). IFM features some integration of the light field into discrete points and dispersing these points spectrally to efficiently fill the detector. There are several implementations of IFM, for instance, using fibre optic arrays [16] and microlens arrays [23], with the latter offering a particularly compact, facile and low-loss solution.

Here, we demonstrate the use of snapshot hyperspectral imaging based on integral field mapping using a microlens array for rapid, widefield, spectrally resolved mapping of intracellular laser emission. Our study focuses on disk-shaped whispering gallery mode (WGM) lasers, made from a III/V semiconductor multi-quantum well material [3,24]. The high refractive index of these lasers allows for a sub-micron cavity size (total volume, ~0.1 μm$^3$), much smaller than the nucleus of a eukaryotic cell (~100 μm$^3$), which is essential for preserving normal cell function. In contrast to point scanning, our SSHI method enables spatial and spectral detection in a single acquisition, which increases the imaging speed from minutes [4,9] to video rate. Furthermore, for low-light samples, the technique can be much faster since the laser emission can be collected concurrently over long integration times instead of sequentially for each individual location.

We calibrate our system for the detection of the narrow linewidths of the disk lasers and discuss the opportunities and challenges of snapshot detection schemes in this area. We experimentally demonstrate a spatial resolution of 5 μm and a spectral resolution of under 0.8 nm. We demonstrate the utility of our system for volumetric hyperspectral imaging via objective scanning, achieving a 250 × 250 × 800 μm$^3$ volume size in $x$, $y$ and $z$, respectively. We further demonstrate widefield hyperspectral detection over a 3 × 3 mm$^2$ field of view in a few minutes by sequentially tiling 17 × 17 snapshot regions. The imaging speed was limited by the speed of the motorized stages and the integration time required to achieve a sufficient signal to noise ratio. Within each field of view, our technique is inertia-free, i.e., it does not require beam scanning, and thus is relatively simple to control without the need for expensive data acquisition cards. It can be seamlessly integrated into existing microscopes, making it a versatile and easy-to-use imaging solution. This detection scheme offers new opportunities for high-throughput, widefield and volumetric imaging, and is compatible with widefield excitation schemes and reduces photodamage.

## 2. Methods

### 2.1 Experimental setup

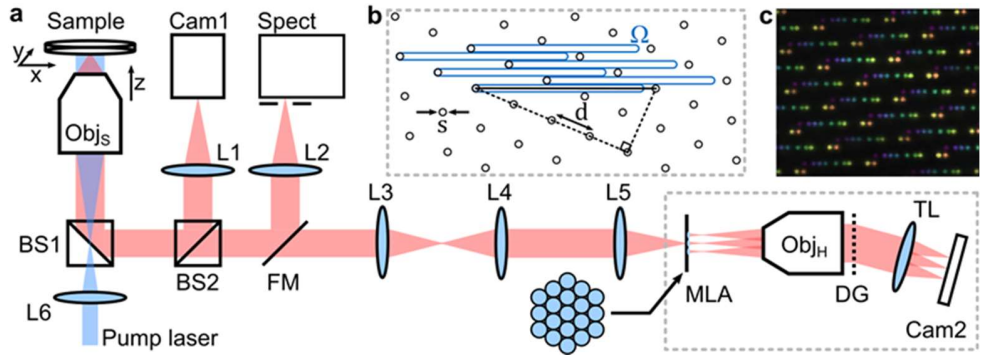

Fig. 1. Optical setup. (a) Optical layout of the snapshot hyperspectral imaging system, including widefield brightfield and fluorescence imaging and conventional spectrometer for reference measurements. Obj$_S$ and Obj$_H$: sample and hyperspectral objectives; L1–6: lenses; BS1–2: beam splitters; FM: flip mirror; MLA: microlens array; DG: diffraction grating; TL: tube lens (telemacro). (b) Illustration of the field mapping via the spectral dispersion of the hexagonal array. (c) Experimental image detected on the hyperspectral camera coloured using false colours based on wavelength.

Figure 1 shows the experimental setup developed in this work. A custom-built inverted microscope (Cerna modular pieces, Thorlabs, USA) is equipped with an objective lens (Obj$_S$, Plan Apochromat, 20x 0.75 NA, Nikon, Japan) mounted on a motorised focusing module (ZFM2020, Thorlabs, USA). The sample is mounted on a two-axis translation stage (PLS-XY, Thorlabs, USA), equipped with an on-stage incubator (H301, Okolab, Italy) for cell measurements, set at 37°C and purged with a 5% $CO_2$:air mixture. For brightfield measurements, the sample is illuminated with a white LED and Koehler illumination optics in transmission. A tunable pulsed optical parametric oscillator laser system (OPO, 5 ns, 20 Hz pulse, OPOTEK, USA), is used for the calibration of the hyperspectral imaging, and a 473 nm pulsed blue diode laser (1.5 ns, 1 kHz pulse, Alphalas, Germany) is used to pump the disk lasers. A lens focuses the pump light to the back focal plane of the objective, such that the light arrives at the sample plane collimated. The emitted and reflected light is coupled to the collection arm via a beamsplitter (BS1). The sample image is relayed to the SSHI unit via a 4f lens system (L4, L5). To provide context to the hyperspectral imaging, a removable beamsplitter (BS2: 10R/90T) can be used to image the brightfield and lasing intensity of the sample with a widefield camera (Cam1, Retiga EXi CCD, QImaging, Canada). An additional flip mirror allows point measurements of the lasing signal from the centre of the sample using a spectrometer equipped with a CCD detector (SR500i, Andor, UK) for calibration and validation purposes. In-house software based on MATLAB is used to synchronise the light delivery, camera acquisitions, and $x$, $y$ and $z$ scanning.

### 2.2 Snapshot hyperspectral imaging

Our embodiment of the IFM SSHI system, which is inspired by the geometry in Boniface et al. [25], is highlighted in Fig. 1(a) by the grey dashed box. A microlens array (MLA, 18-00079, SUSS MicroOptics, Switzerland) placed at the image plane integrates the detected light field into an evenly distributed hexagonal array of foci. This array of foci is re-imaged by a secondary objective (ObjH, Plan N 10x 0.25 NA, Olympus, Japan). A diffraction grating (DG, 300 grooves/mm, GT13-03, Thorlabs, USA) placed at the back aperture of the objective spectrally disperses the light, which is then imaged by the camera (Cam2, Orca Flash 4.0, Hamamatsu, Japan) using a tube lens (TL). This results in a hexagonal point array that is spatially dispersed if the incident light is composed of different wavelengths (Figs. 1(b) and (c)). The rotation angle of the diffraction grating with respect to the microlens array is chosen to optimally fill the camera [23]. We utilised a commercially available telemacro lens as the tube lens (TL; Tele-Macro AF 70-300mm, Tamron, Germany) to reduce the footprint of the SSHI unit and provide the capacity to tune the magnification.

The pitch (spacing), $d$, and numerical aperture, NA, of the microlenses determine the maximal spectral sampling. The size of the foci of the microlens array, $s$, is determined by the Abbe diffraction limit, $s = 0.61\lambda/\text{NA}$ (2.6 μm for our system). Based on the angle of dispersion with respect to the hexagonal lattice illustrated in Fig. 1 (b), the separation between microlenses along the dispersion axis is $\sqrt{19}d$, and thus the number of resolvable spectra is: $N_\lambda = \sqrt{19}d/s$. The spectral bandwidth, $\Omega$, dispersed by the diffraction grating with a dispersion angle, $d\beta/d\lambda$, must fit between the microlenses. The total dispersion at the camera is $\Omega\, d\beta/d\lambda\, f_\text{TL}$, which when adjusted for the camera magnification, $M_C = f_\text{TL}/f_{\text{Obj}_\text{H}}$, must satisfy: $\Omega < \sqrt{19}d(f_{\text{Obj}_\text{H}} d\beta/d\lambda)^{-1}$. This results in a spectral resolution: $r_\lambda = \Omega/N_\lambda$.

The spatial FOV and resolution are determined by the magnification, $M$, of the sample image onto the microlens array and the total pixels available in the camera, $P$. The spatial resolution is: $r_x = d/M$, and the spatial sampling per FOV is: $N_x = P/(d'\sqrt{3}/2)$, where $d'$ is the microlens pitch on the camera in pixels: $d' = df_\text{TL}/(pf_{\text{Obj}_\text{H}})$; $p$ is the physical pixel size.

Optimising our system for hyperspectral imaging of disk laser emission, we used a microlens array with a pitch $d$ of 30 μm and NA of 0.16, an objective lens with an $f_{\text{Obj}_\text{H}}$ of 18 mm, a diffraction grating with a $d\beta/d\lambda$ of $0.31 \times 10^{-6}$ rad/m, and a camera with $2048 \times 2048$ pixels. At a central wavelength of 670 nm, this resulted in a theoretical spectral bandwidth, $\Omega$, of 24 nm, comparable to the emission bandwidth of our microdisk lasers [3], and a spectral resolution, $r_\lambda$, of 0.46 nm, leading to $N_\lambda = 51$ resolvable spectra. Ultimately the speed of the acquisition of large areas is limited by the desired spatial resolution. With this trade-off in mind and given the imaging requirements, such as the typical volume of a mammalian cell ($\sim 10^3$ μm$^3$) and diameter of the microdisk lasers (1-3 μm), we chose a magnification of $M = 6$ to achieve a single cell spatial resolution of 5 μm. The resolution was set by the effective microlens pitch at the sample, which was verified experimentally using a ruled test target. The variable $f_\text{TL}$ telephoto lens was tuned to fit the full aperture of the microlens array, which has a spatial sampling of $N_x = 50$, leading to a FOV of $250 \times 250$ μm$^2$.

### 2.3 Hyperspectral image calibration and reconstruction

The reconstruction of hypercubes from hyperspectral snapshots requires the precise mapping of spatial and spectral sample intensities from spatial camera coordinates. A hexagonal lattice can be described using two basis vectors, $\mathbf{e_1}$ and $\mathbf{e_2}$, and an offset, $\mathbf{x_0}$, such that the position of each spot, $\mathbf{x_i}$, is within the set: $\{\mathbf{x} = n\mathbf{e_1} + m\mathbf{e_2} + \mathbf{x_0} \mid n, m \in \mathbb{Z}\ \text{and}\ \mathbf{x} \in C\}$, where $C$ denotes the spatial bounds of the camera. The basis vectors describe two independent vectors of separation between adjacent foci, and the offset describes the position vector of a spot at the central wavelength, $\lambda_0$. The hexagonal lattice positions at a wavelength, $\lambda$, can then be described as: $\mathbf{x}_\lambda = \mathbf{x} + \alpha(\lambda - \lambda_0)\hat{\mathbf{e}}_\mathbf{d}$, where $\alpha = dx/d\lambda$ is the rate of dispersion in pixels per wavelength, and $\hat{\mathbf{e}}_\mathbf{d}$ is the unit vector along the dispersion axis.

As such, the calibration of our SSHI system requires accurate measurement of the basis vectors, offset vector, and the dispersion. To do so, we perform calibration by illuminating a scattering sample with several wavelengths using a tunable laser within the bandwidth of detection. Each wavelength generates a hexagonal lattice in the snapshot image. The basis vectors and offset can be readily measured from a single calibration image using one wavelength. Using at least two wavelengths, the dispersion parameter can also be readily estimated from the relative shift in the hexagonal lattice. We further image a micrometre ruler with both brightfield and SSHI to aid with the co-registration of the camera images. This procedure was performed prior to each set of experiments.

After the calibration is performed, raster hypercubes, $I(x, y, \lambda)$, are reconstructed by using scattered interpolation of the snapshot intensities queried at the lattice coordinates, $\mathbf{x}_\lambda$, over a regular grid, for each sample wavelength. Specifically, we implement the native scattered interpolation method in MATLAB, which is based on Delaunay triangulation [26]. We generate hypercubes with dimensions of $100 \times 100 \times 100$ pixels, in $x$, $y$ and $\lambda$ to satisfy the Nyquist sampling criterion. Smoothing the snapshot intensity image in the dispersion axis by the spectral sampling size and in the orthogonal axis by the separation distance between adjacent spectral lines further ensured that information was not lost during image reconstruction. Hyperspectral reconstruction took less than 1 s per snapshot. The calibration and hyperspectral reconstruction code is provided as open source, as detailed in the data availability statement.

### 2.4 Fabrication of disk lasers

The fabrication of disk-shaped WGM microlasers for this study largely followed our previously reported protocols [3,24]. In brief, we used a heterostructure grown on GaAs wafers and made up of layers of InGaP wells and AlInGaP barriers, forming a double quantum well structure with a total thickness of 180 nm that is located on an AlGaAs sacrificial layer as described in detail in [3,24,27] (EPSRC National Epitaxy Facility, Sheffield, UK). Substrates were cleaned by a 3 min sonication in isopropanol, acetone, deionised water and methanol, followed by 3 min of $O_2$ plasma. The substrates were spin-coated with SU8 photoresist (SU8 2000.5, KayakuAM, USA; 3:1 dilution with cyclopentanone) and soft-baked on a hotplate at 90°C for 2 min. After cooling down, the sample was exposed with a UV mask aligner using a custom mask with 3-µm diameter holes, followed by a post-exposure bake for 2 min at 90°C. The photoresist was developed in 2-methoxy-1-methylethyl acetate (EC solvent, Microposit, Germany) for 60 s and then cured at 180°C for 5 min. A 30 s plasma descumming step was performed, followed by a 12 s wet etch in aqueous solution of HBr (1 M) and $Br_2$ (0.4 M), which defined the circular disk shape. The SU8 photoresist caps were removed by reactive ion etching in $O_2$ plasma for 7 min. A subsequent selective 3 min wet etch in 5% HF collapsed the disk onto the GaAs substrate. The resulting disk diameter was 1–3 µm and the thickness of the disks was 180 nm.

### 2.5 Cell culture and disk internalisation

Macrophage cells were isolated from blood samples obtained from healthy human donors after ethical review (School of Medicine, University of St Andrews) and under informed ethical consent. They were cultured in RPMI supplemented with 10% fetal bovine serum and 1% penicillin-streptomyocin in ibidi dishes (µ-dishes, Ibidi, Germany). Wafer pieces containing disk lasers were incubated in isopropanol for several minutes to ensure sterility. They were washed in PBS and finally harvested in a 2 ml tube by sonication directly into cell culture medium. The solution was filtered through a 5 µm pore size to remove any fragments of wafer and added to the ibidi dishes with adherent cells. The cells were allowed to uptake disks by naturally occurring phagocytosis overnight before the measurements.

*2.6 Phasor analysis*

For visualisation of widefield hyperspectral data, we utilise an approach inspired by phasor analysis in fluorescence imaging [28]. The phasor approach is a fit-free method of obtaining the centre of mass (phasor angle) and the spectral width (phasor amplitude) of the emission spectrum, and a convenient method for visualising multidimensional information. We perform a single normalised discrete Fourier transform for each spectrum:

$$C = \sum_{\lambda \in \Omega} I_\lambda \cdot e^{-i2\pi\lambda/\Omega} \Big/ \sum_{\lambda \in \Omega} I_\lambda.$$

From this, we can readily extract several useful parameters, including the lasing peak centre of mass, $\angle C$, where the angle in $[0, 2\pi]$ corresponds linearly to the wavelength in $\Omega$; and spectral width, $||C||_2$, which is in $[0, 1]$ and corresponds to the relative width of the peak (where 0 is a uniform distribution and 1 is a delta function).

## 3. Results

*3.1 Calibration*

We first calibrated the performance of our hyperspectral imaging system using a tunable optical parametric oscillator (OPO) laser system and a reference spectrometer. This was accomplished by illuminating a scattering sample sequentially with two distinct wavelengths within the calculated spectral bandwidth of our system, recording the scattered light by both the SSHI unit and the reference spectrometer, and finally determining the basis vectors, offset, and spectral dispersion parameters from this data (see Methods). We then validated the calibration by recording a set of additional wavelengths. Figure 2 shows the normalised spectra from nine distinct illumination wavelengths recorded using SSHI, which compares favourably with the data obtained by the reference spectrometer. The peak position and width of each spectral peak were estimated using a Voigt fit [13]. The resulting spectral peak positions matched well with the ground truth spectra, with a mean absolute error of 7 pm. This very close agreement indicates that the spectral dispersion in the SSHI is linear and does not require more than two wavelengths for calibration. The mean full-width at half-maximum (FWHM) of the spectra acquired by SSHI was 0.51 nm, which is slightly larger than the expected spectral resolution of 0.46 nm, likely due to the spherical aberration by the microlenses leading to a larger-than-ideal spot size.

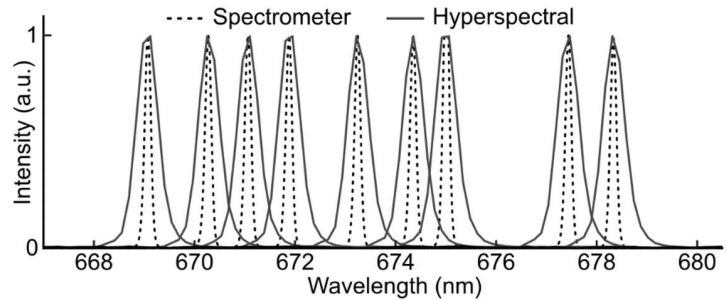

Fig. 2. Spectral calibration of hyperspectral imaging using a reference spectrometer and a tunable laser system. Mean normalised spectral response recorded by SSHI unit and by the reference spectrometer.

*3.2 Hyperspectral imaging of disk lasers*

Next, we evaluate the capacity of the calibrated SSHI system to image disk lasers. Figure 3 shows the lasing spectra of representative disk lasers in PBS solution reconstructed using SSHI unit compared with the ground truth spectra. Figure 3 also shows the maximum intensity projection of all wavelengths in the recorded hypercube. We find a good agreement between the individual disk laser spectra with the ground truth spectra, with a 0.52 nm mean absolute error in peak position for the six lasers depicted in Fig. 3. Fig 3(c) and (d) demonstrate that our SSHI system can record spectra from more than one laser in the FOV from a single snapshot. By contrast, the spectrometer reference measurements required the disk lasers to be addressed individually by translating them to the centre of the FOV. We further note that relative differences in the spectral intensities of different lasers were reproduced in a similar manner by SSHI and the spectrometer reference measurement.

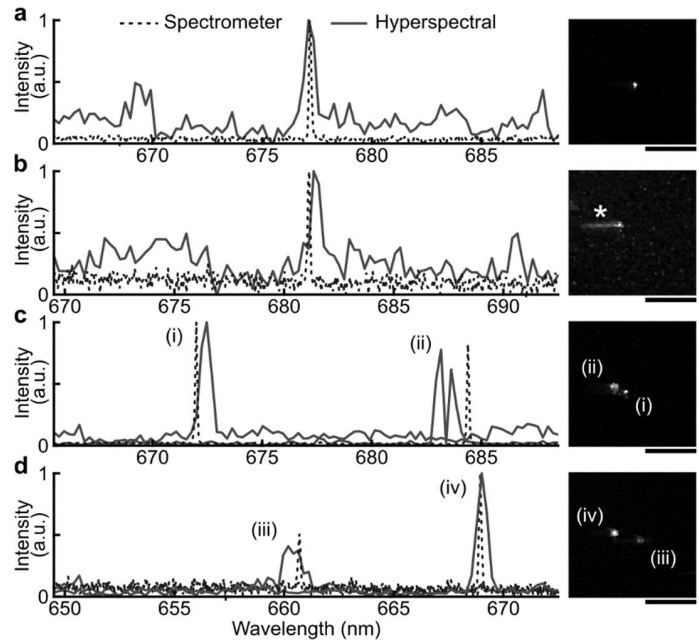

Fig. 3. Hyperspectral imaging of microdisk lasers. Laser spectra reconstructed from hyperspectral data and spectrometer reference measurements (left) and maximum spectral intensity projections (right). (a) FOV with single laser operating well above its lasing threshold. (b) FOV with a single laser showing fluorescence background emission (marked with *) in addition to the lasing peak. (c) and (d) FOVs with multiple lasers (labelled i to iv). Scale bars denote 100 μm.

Figure 3 also illustrates different cases of SNR in the reconstructed spectrum. While the noise floor is well visible in the spectra in Fig. 3(a,b), spectra in Fig. 3(c,d) show excellent SNR. The SNR of our measurement depends on the brightness of the disk laser and the optical collection efficiency. An intrinsic characteristic of the WGM disk lasers is that their emission is predominantly in plane [29,30]. Consequently, the collection efficiency is influenced by the tilt angle of the disk laser relative to the collection numerical aperture of the objective [30]. The unique emission properties of the disk lasers coupled with the SSHI detection can give rise to several artefacts, which are illustrated by the measurements in Fig. 3 and which we discuss in turn. In Fig. 3(b), there is a tail in the intensity image on the right, corresponding to the fluorescence of the disk laser. This can also be verified by the elevated background in the corresponding spectrometer measurement. Fluorescence can be present when disks lasers are close to the lasing threshold. Because of the broadband spectral emission outside of the design bandwidth of the SSHI unit, fluorescence is incorrectly mapped to adjacent pixels. For sparsely distributed disk lasers, this can be trivially filtered from the image intensity or spectral peak width. To further improve the situation, an optical filter commensurate with the calibration bandwidth of the SSHI unit could be added to the detection path.

Defocusing and scattering effects are also evident from the intensity images and the spectra. The disk lasers are smaller than the nominal spatial resolution of our system (i.e., < 5 µm) and, thus, should appear as point sources. Scattering of the laser emission in the media leads to non-uniform spatial broadening of the laser intensity, which is further emphasised by the anisotropic emission of disk lasers [29,30]. This scattered light is defocused and not conjugated to the microlens array. For a uniform defocus, we expect a larger spot size after the microlens array, leading to spectral broadening. A non-uniform intensity and phase profile within the microlens integration area (5 µm) may lead to a non-paraxial incidence on the diffraction grating and, thus, a spectral shift. In fact, such spatial shifting due to the integration of non-uniform wavefronts is the mechanism of Shack-Hartmann wavefront sensors [31]. Here, scattering and non-uniform intensity have likely contributed to the observed lower than expected spectral resolution and to the deviations between the measured peak positions and the nominal laser wavelengths (as determined by the spectrometer). Specifically, the mean FWHM for the lasers shown in Fig. 3 is 0.79 nm for SSHI and 0.12 nm for the spectrometer measurements, while the mean error in peak position is 0.52 nm (measured relative to the spectrometer measurement). The spectrometer resolution was limited by the slit size and grating dispersion, chosen to ensure sufficient light collection and, thus, signal-to-noise ratio (SNR). We observed that the disks with fewer artefacts displayed a nominal FWHM similar to the resolution achieved during the calibration measurements, i.e., 0.52 nm. For example, the FWHM of the disk laser in Fig. 3(a) is 0.59 nm, and disk laser (i) in Fig. 3(c) even has a FWHM of 0.45 nm. An extreme negative example, which is associated with the distorted profile in the corresponding intensity image, is disk laser (ii) in Fig 3(c) where we observe a dual emission peak and a combined FWHM over both peaks of 1.34 nm. We will discuss strategies to overcome the artefacts described above and the associated challenges in the Discussion section.

### 3.3 Reproducibility in translated disk lasers

Because of the potential for artefacts from scattering and defocussed light, we also evaluate the reproducibility of spectral reconstruction from disk lasers translated across the FOV. First, we evaluate how a non-uniform intensity incident on a microlens affects spectral accuracy. We do this by translating a single disk laser in 1 µm steps over 10 µm to make sure that the laser emission traverses the receptive field of several microlenses. Figures 4(a–c) show the disk laser position with respect to the hexagonal array and the reconstructed spatial and spectral intensities. At the centre of the disk laser, the peak positions are relatively consistent with a 75 pm mean absolute error compared to the spectrometer reference. The inset in Fig. 4(a) shows a small linear shift in the spot intensity as the laser moves relative to the microlens.

Figures 4(a) and (b) also illustrate the effect of scattering in the raw and reconstructed images (marked by an asterisk in each image). The microlens pitch at the image plane is 5 µm, which exceeds the disk laser diameter of 1–3 µm. This indicates any light collected away from the microlenses conjugate to the disk laser position must correspond to scattering or defocussed imaging. The spectral broadening and dispersion resulting from the scattering is evident in Fig. 4(b) by the broad position-dependent blurring in the raw image obtained with the SSHI unit, which then leads to spectral shifts away from the centre of the disk laser in the reconstructed image (Fig. 4(b)). However, at the position of the disk laser, i.e., the position corresponding to the highest intensity, the spectra closely match that of the reference spectrometer.

Figures 4(d) and (e) illustrate the capacity to track a single disk laser undergoing large spatial shifts, which, for instance, would be encountered in tracking, dynamic and microfluidic imaging applications. The disk lasers were translated by 50–100 µm using a motorised microscope translation stage (PLS-XY, Thorlabs, USA) and a hand-operated controller (MCM3001, Thorlabs, USA). The intensity images in Fig. 4(d) overlay the maximum peak intensity of the same disk laser from the entire sequence of positions. The spectra shown were obtained from the centre of the disk laser at each position, i.e., the position of highest intensity.

The spectral peak positions correspond well to the spectrometer measurements, which were taken at location (1) in each case. The lasing intensity is not uniform across the FOV, but this is at least in part due to a spatial variation in the pump laser intensity. Additional peaks are observed for some positions (e.g., position 3, Fig. 4(e)). This is likely due to the presence of increased scattering. These variations have likely arisen from the minute shifts in disk laser position and scatterer rearrangement due to the inertia of translation. Despite this, the main peak corresponding to the unscattered laser emission remains prominent in the spectrum. The mean absolute error in peak position was 0.089 nm and 0.21 nm for Figs. 4(d) and (e) respectively.

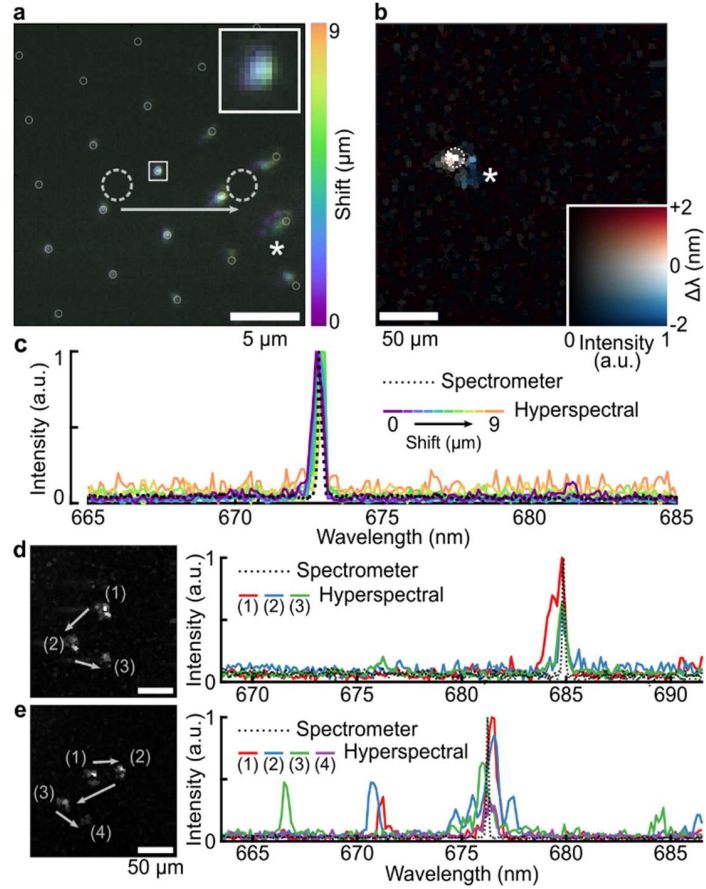

Fig. 4. Influence of laser position on spectral detection. (a–c) Hyperspectral detection of a disk laser translated in 1 μm steps over 10 μm. (a) Raw hyperspectral images with each position coded by false colour. Solid circles correspond to the expected spot positions at the spectrometer reference wavelength. Dashed circles correspond to the laser start and finish position. The inset shows the spectral peak at the central microlens. (b) Hyperspectral image with maximum peak intensity in grey and the difference between peak wavelength determined by SSHI and the spectrometer reference in false colour. Inset shows the two-dimensional colour scale bar. Asterisks in (a, b) illustrate the effect of scattering. (c) Spectra of the disk laser during the 10 μm translation performed in a. (d, e) Data for translation of disk lasers over 50-100 μm, showing maximum peak intensity overlaid for all positions (marked as 1–4, left) and the corresponding spectra obtained at the centre position in each frame (right).

### 3.4 Volumetric hyperspectral imaging

Next, we demonstrate how our SSHI method can perform volumetric hyperspectral imaging. A volumetric sample was prepared by suspending disk lasers in agarose gel. For this, disk lasers where mixed with a heated aqueous solution of 5% wt agarose and pipetted into a glass-bottom dish (μ-dishes, Ibidi, Germany). The agarose was rapidly cooled in a freezer (-20°C) to ensure disk lasers remained suspended. Figure 5 visualises the hyperspectral information in a FOV containing three disk lasers. The volume was recorded in the region of interest by scanning the objective lens with an automated focusing module (ZFM2020, Thorlabs, USA). The volume comprised 160 sequential scans with a 5 μm $z$ separation, leading to a 250 × 250 × 800 μm$^3$ total volume size in $x$, $y$ and $z$, respectively. The volume of the peak spectral intensity is visualised in Fig. 5(a). A linear transparency map applied to the normalised intensity enables the visualisation of the recorded conical scattering profile of the three lasers (marked 1–3). The conical scattering profile is a product of the numerical aperture of the objective and the anisotropic in-plane disk laser emission [29,30].

The microlenses in the SSHI unit acts as individual apertures, conjugated to the imaging plane. The light field from defocused regions will be not only spread across several microlenses, but also will possess a non-planar wavefront. This leads to not only blurring in the spatial intensity, such as would be encountered in brightfield microscopy, but additional blurring and spatial shift on the hyperspectral detector. This manifests as spectral blurring and shifting, which, to a large extent, is filtered out by the image field mapping (similarly to what is illustrated in Fig. 4(a)). As evident in Fig. 5, the rejection of out-of-plane light in the reconstructed lasing intensity enables optical sectioning in SSHI. As a consequence, we can localise the laser position in a volume, and determine its absolute position, which is not possible in a conventional widefield imaging system. The distances between the disk lasers determined in this way are given in Fig. 5(a). Figure 5(b) shows the spectra from the centre of each of the three disk lasers. We see a sharp spectral peak and few artefacts from scattering, indicating that when lasers are positioned away from optical interfaces, SSHI can provide high quality spectra.

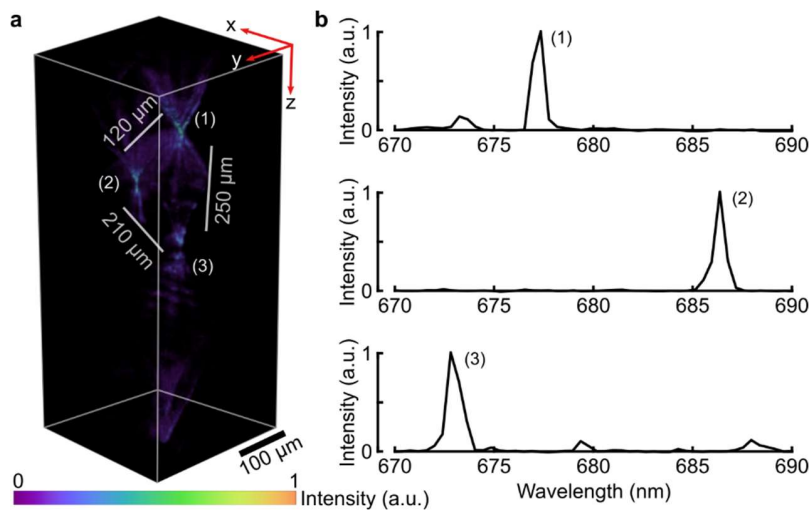

Fig. 5. Volumetric imaging of an agarose phantom loaded with disk lasers obtained by SSHI. (a) Volumetric rendering of the peak spectral intensity at each point using a false colour scale and linear transparency. Three separate disk lasers, numbered 1-3, can be clearly identified; their respective distances are given in the image. (b) Corresponding spectra at the centre position of each laser.

## 3.5 Widefield hyperspectral imaging in biology

Finally, we showcase the ability to perform widefield hyperspectral imaging in biologically relevant settings by imaging disk lasers internalised by macrophage cells. To maximize the investigated area, we sequentially tile 17 × 17 snapshot acquisitions using an automated $xy$ translation stage. As a result, we can obtain a panoramic view of the cells and of individual disk laser spectra with an effective total FOV measuring 3,240 × 3,240 µm², as shown in Fig. 6(a). Brightfield, lasing emission and SSHI were taken at each of the 289 snapshot positions, which required less than 9 min in total.

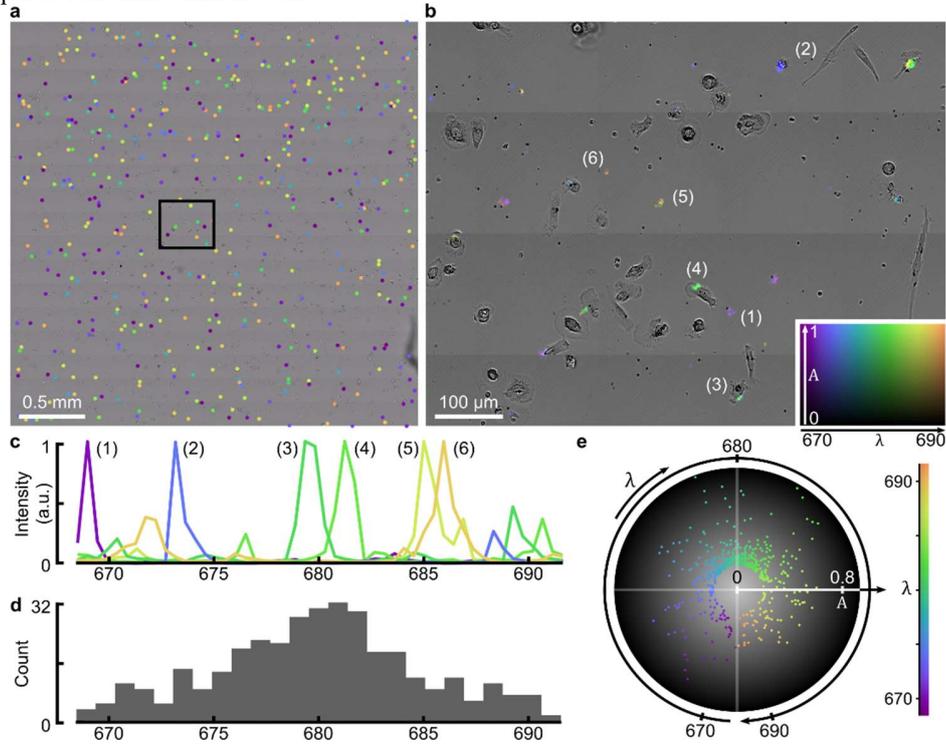

Fig. 6. Widefield hyperspectral imaging of cells containing intracellular disk lasers. (a) Position and central wavelength of the brightest 400 disk lasers overlaid on a brightfield image of the culture of macrophages on which the measurement is performed. (b) Amplitude (A) and peak wavelength (λ) of lasers across the zoomed-in region marked by a black square in (a). False colour representation as per the colour scale in the lower right inset. (c) Disk laser spectra from locations marked as (1–6) in (b). (d) Histogram of the central wavelength of all lasers in the field of view shown in a. (e) Phasor plot distribution of all laser spectra. Amplitude (A) is the sharpness of the emission peak, and the angle (λ) is the central wavelength of the emission spectra. Scale bars are 0.5 mm in (a) and 100 µm in (b).

Figure 6(a) visualises the spatial position and emission wavelength of 400 brightest disk lasers as points overlaid on the brightfield image. A zoomed-in region, indicated by the black square, is shown in Fig. 6(b). The false colour map indicates both the lasing central wavelength (λ) and the amplitude (A) of the lasing peak, calculated using phasor analysis described in the Methods section [28]. This analysis simplifies the visualisation of large numbers of laser spectra in widefield images. Figure 6(c) shows representative spectra from six disk lasers marked in Fig. 6(b). The distribution of the peak lasing wavelength for all of the 400 lasers is shown in Figs. 6(d) and (e), with the histogram in Fig. 6(d) validating that the designed disk laser emission range is commensurate with the selected bandwidth of the SSHI system. The phasor plot in Fig. 6(e) displays the individual lasers using their phasor amplitude (A), which represents the sharpness of the spectral emission peak and, in disk lasers, is proportional to the laser brightness and the SNR of the detection. The angle of the phasor plot in Fig. 6(e) describes the central wavelength of the laser emission. These results highlight the promise of rapid SSHI of disk lasers for multiplexed sensing.

## 4. Discussion

Snapshot hyperspectral imaging promises rapid widefield readout of the spectra emitted by intracellular lasers. We demonstrated an embodiment of SSHI using a microlens array that can measure and distinguish disk lasers based on their spatial and spectral intensity and peak wavelength over large FOV and in 3D. The development of such methods can unveil new applications for multiplexed rapid sensing of biological properties with cellular precision.

The acquisition rate of our SSHI system is given by the integration time of the camera, which in the present case was set to 100 ms to ensure a sufficient SNR is reached, leading to a maximum imaging rate of 10 hypercubes per second. We expect that in the future the integration time can be shortened by at least 10-fold by optimising light throughput, repetition rate of the pump laser, camera sensitivity and the brightness of the used intracellular lasers. Even for the relatively low camera frame rate used in the present study, the equivalent acquisition time when using a raster scanning approach across the same number of pixels per FOV (100 × 100 pixels) is 10 μs per sampled spectrum. Fast linescan cameras can in principle support raster-scanning spectral acquisition at such rates; sub 1-ms spectral acquisition times have already been demonstrated using an InGaAs linescan camera to record the spectra emitted by disk lasers [4]. However, this requires collecting a relatively large amount of light from a small volume in the sample at a time, which in turn requires high intensity excitation light and thus can lead to issues with phototoxicity and photodamage to the intracellular lasers. Instead, widefield detection methods, such as the SSHI developed in this study, make use of all light emitted across the FOV, and thus we expect that they ultimately enable faster detection and lower photodamage than raster scanning approaches.

The major challenge in SSHI methods is the necessary trade-off in the spatial and spectral resolution. In an IFM embodiment of SSHI that uses a microlens array, the resolution and FOV/bandwidth of the spatial and spectral detection can be readily tuned by varying magnification and dispersion properties. However, the need to provide both high spatial and spectral resolution with the same camera, coupled with the spectral broadening arising from scattering and defocus artefacts, makes the use of SSHI challenging in applications that rely on detecting minute spectral mode shifts of intracellular lasers. Instead, SSHI methods could excel in wide area multiplexing, such as for single cell tracking in 3D tissue and high-throughput applications such as microfluidics [13].

While SSHI has been the subject of several studies in the past [16], its use to record the spectra emitted by intracellular lasers is new and presents several specific challenges. For instance, in the field of intracellular lasing, an emphasis is placed on laser miniaturisation to ensure minimal impact on cells and tissue [32]. Thus, when evaluating SSHI, it is pertinent to consider disk lasers as sub-diffraction-limited point sources. Conventional SSHI methods have been designed to sample scenes with the expected power spectral content of natural images, i.e., a logarithmic attenuation of intensity with spatial frequency [33]. However, the imaged light field from point-sources, like narrow-linewidth disk lasers, comprises exceptionally high spatio-spectral frequencies. As a consequence, the light field incident on each microlens is no longer approximated by a plane wave, leading to imperfect field integration. Practically, in our study, this has manifested in the observed artefacts from scattering and defocus.

Early implementations of SSHI have flourished in the field of astronomy [16]. Many such methods were designed to record distant objects. The incoming near-planar wavefronts enabled ready field integration using optical elements with a proportionally higher etendue. This requirement for higher etendue in the integration optics compared to the detection optics has been noted in IFM SSHI [16]. Microscopy, however, is characterised by the use of high-NA imaging optics. Microlenses are unlikely to match the NA of typical microscopy objectives, and higher NA microlenses present substantial challenges in alignment and stability. This presents a major challenge in overcoming field inhomogeneity and the resulting scattering and defocus artefacts. Instead, using lower-NA objectives, commensurate with the reduced spatial resolution of SSHI, would overcome this challenge, albeit at the cost of lower collection efficiency. Alternatively, spatial filtering might be used, for instance, by using a matched pinhole array at the foci of the microlenses or by using coherent fibre bundle IFM [16]; however, these methods introduce a high loss in the detected light intensity.

Improvements and modification in disk lasers fabrication, such as modifications to the cavity geometry [34] and introduction of defects, can lead to omnidirectional emission [30]. Using lasers with such modifications could improve the speed and the accuracy of SSHI detection as their isotropic omnidirectional emission would prevent a situation where scattered in-plane emission from lasers overshadows the direct signal from a disk laser located in focus. It is important to note, however, that we found scattering artefacts to be most prominent in samples where disks were located on a glass substrate (in Figs 3 and 4), and less noticeable in volumetric and in-cell measurements (in Figs 5 and 6).

SSHI can be further enhanced using digital and computational imaging approaches [35]. The rejection of defocused light and, thus, defocusing artefacts, may be achieved through digital pinholes. For instance, a similar effect to confocal pinholes can be achieved via a spatio-temporal modulation of the image intensity and, subsequently, digital demodulation [36]. This can be realised using a physical coded aperture or a digital micromirror device [37,38]. Coded aperture-based encoding can also be used in the spatio-spectral domain to realise CASSI [22]. CASSI enables compressive sensing of hypercubes with a strong potential for disk laser sensing. Exploiting the natural spatial and spectral sparsity of disk laser emission has the potential to substantially mitigate the trade-offs in resolution. However, the reliance on solving an underdetermined problem and the challenges in physical implementation could lead to new errors and artefacts in quantitative sensing [16].

## 5. Conclusions

We demonstrated rapid snapshot hyperspectral imaging of intracellular lasers, implemented using integral field mapping with a microlens array. We characterised the performance of the system in detecting distinct emission spectra from micron-sized disk lasers, demonstrating a spatial resolution of 5 μm and spectral resolution of under 0.8 nm. We then applied this method to widefield imaging in cells over $3 \times 3$ mm$^2$ areas and to rapid volumetric imaging to depths of 800 μm. The unique geometry and emission spectra of intracellular lasers make it a distinct and interesting challenge for SSHI methods. We show that while SSHI systems must be carefully tuned to intracellular laser detection, they offer new opportunities towards high-throughput and massively multiplexed detection of disk lasers and high-throughput precision biosensing applications.

**Acknowledgements**

We thank Dr Ivan Gusachenko for designing the initial microlens array spectrometer and Dr Simon J. Powis for the isolation of human microphages.

**Data availability**

The reconstruction code for SSHI is available at https://github.com/philipwijesinghe/snapshot-hyperspectal-imaging. The data underpinning this work is available at https://doi.org/10.17630/4fe7ae1c-9e0d-4671-a795-f29cb64d504c [39].

**Funding**

This work received financial support from a UK EPSRC Programme Grant (EP/P030017/1). PW was supported by the 1851 Research Fellowship from the Royal Commission. KD acknowledges support from the Australian Research Council (FL210100099). MCG acknowledges support from the Alexander von Humboldt Foundation (Humboldt professorship).